\documentclass[useAMS,usenatbib]{mnras}

\usepackage{graphicx,amssymb,hyperref}
\usepackage[fleqn]{amsmath}  
\usepackage{multirow}
\usepackage{graphicx} 
\usepackage{color}

\usepackage{graphicx,amssymb,multirow}
\usepackage[fleqn]{amsmath}
\usepackage{abbrev}

\def\mnras{MNRAS}
\def\apj{ApJ}

\def\simlt{\lower.5ex\hbox{$\; \buildrel < \over \sim \;$}}
\def\simgt{\lower.5ex\hbox{$\; \buildrel > \over \sim \;$}}
\def\etal{{\it et al.}}

\def\mrho{\langle\rho\rangle}

\def\rmd{\mathrm{d}}
\def\rv{r_{\mathrm{v}}}
\def\rmm{\mathrm{M}}

\newcommand{\be}{\begin{equation}}
\newcommand{\ee}{\end{equation}}
\newcommand{\ba}{\begin{eqnarray}}
\newcommand{\ea}{\end{eqnarray}}

\usepackage{graphicx}

\usepackage{nicefrac}
\usepackage{units}
\usepackage{flushend}

\DeclareGraphicsExtensions{.pdf,.png}

\title[Dark matter scattering rates]{Self-interacting dark matter scattering rates through cosmic time}

\author[A.\ Robertson \etal]{Andrew Robertson\thanks{e-mail: {\tt andrew.robertson@durham.ac.uk}}, 
Richard Massey,
Vincent Eke,
Richard Bower\\Institute for Computational Cosmology, Durham University, South Road, Durham DH1 3LE, UK\\
}

\begin{document}
\date{ Accepted ---. Received ---; in original form \today.}
\pagerange{\pageref{firstpage}--\pageref{lastpage}} \pubyear{2015}

\maketitle

\label{firstpage}

\begin{abstract}
We estimate the rate of dark matter scattering in collapsed structures throughout the history of the Universe. If the scattering cross-section is velocity-independent, then the canonical picture is correct that scatterings occur mainly at late times. The scattering rate peaks slightly at redshift $z\sim6$, and remains significant today. Half the scatterings occur after $z\sim1$, in structures more massive than $10^{12} \, \mathrm{M}_\odot$. Within a factor of two, these numbers are robust to changes in the assumed astrophysics, and the scatterings would be captured in cosmological simulations. However, for particle physics models with a velocity-dependent cross-section (as for Yukawa potential interactions via a massive mediator), the scattering rate peaks before $z \sim 20$, in objects with mass $\simlt10^{4} \, \mathrm{M}_\odot$. These precise values are sensitive to the redshift-dependent mass-concentration relation and the small-scale cutoff in the matter power spectrum. In extreme cases, the qualitative effect of early interactions may be reminiscent of warm dark matter and strongly affect the subsequent growth of structure. However, these scatterings are being missed in existing cosmological simulations with limited mass resolution.
\end{abstract}


\section{Introduction}

The cold dark matter and cosmological constant model ($\Lambda$CDM) has been successful at describing the observed large-scale structure. However, reported differences on smaller physical scales between simulations and observations \citep[for a review see][]{Weinberg02022015} have raised the exciting question of whether one of the two main assumptions about the dark matter (DM), namely that it is cold with low thermal velocities and that it is collisionless, could need revising. 

Self-Interacting Dark Matter (SIDM) removes the collisionless assumption, such that DM particles have a cross-section for interacting that is sufficient to produce observable astrophysical effects. In the simplest model, originally proposed by \citet{2000PhRvL..84.3760S}, these interactions are elastic scatterings with an interaction cross-section that is independent of velocity. The DM particle collisions decrease the central density of DM haloes and tend to make the DM velocity distribution isotropic, leading to more spherical haloes \citep{2000ApJ...534L.143B,Yoshida:2000gn,2000PhRvL..84.3760S}.

Initial excitement about SIDM was related to its ability to produce constant density cores in dwarf galaxies \citep{Yoshida:2000gn,2000PhRvL..84.3760S}, as well as reduce the amount of substructure in DM haloes.
Constraints placed on the SIDM cross-section imply DM collisions are unlikely to produce significant evaporation of subhaloes, though self interactions can still have a noticeable effect on halo density profiles \citep{Rocha:2013bo}. This could explain the results of a detailed comparison between local dwarf galaxies and simulations made by \citet{2011MNRAS.415L..40B}, who found that the most massive DM substructures around simulated Milky Way-like haloes were considerably more massive than estimated dwarf galaxy masses made from line-of-sight velocity measurements \citep{2009ApJ...704.1274W,2010MNRAS.406.1220W}. If the results from the collisionless $N$-body simulations are representative of the real Universe, then there must be a significant number of massive dark subhaloes around the Milky Way. These subhaloes that do not contain stars despite their large mass have been dubbed `too big to fail'. An alternative explanation is that the most massive subhaloes do form stars, but that their circular velocities are below that seen in the collisionless DM simulations. This can be achieved in SIDM, where the constant density cores formed through DM collisions reduce the circular velocities of subhaloes.

Since SIDM was first proposed as an alternative to collisionless CDM, work has been done to constrain the self-interaction cross-section. Astrophysical considerations have included the core sizes of clusters \citep{Yoshida:2000gn}, the ellipticity of clusters \citep{2002ApJ...564...60M}, evaporation of galaxy haloes in clusters \citep{2001ApJ...561...61G}, and the dynamics and mass-to-light ratios of merging systems such as the Bullet Cluster \citep{Markevitch:2004dl,Randall:2008hs,2015Sci...347.1462H}.

The tightest constraints come from galaxy cluster scales, where the relative velocity between DM particles is high. Meanwhile SIDM's ability to solve the `too big to fail' problem is on the dwarf galaxy scale. This was recently noted by \citet{Fry:2015jx} who found that cross-sections consistent with cluster scale constraints could not significantly reduce the central density of haloes with peak circular velocities below $30 \, \mathrm{km \, s^{-1}}$. For this reason, as well as the fact that many particle physics models give rise to them, there has been increased interest in SIDM with a velocity-dependent cross-section. A DM particle with a cross-section that decreases with increasing relative particle velocity \citep[see e.g.][]{2014JHEP...10..061K} could have an effect on dwarf galaxy scales where velocity dispersions are low, while leaving galaxy clusters relatively untouched. For this reason, we look at a well-motivated particle model that gives rise to a velocity-dependent scattering cross-section.

Assessing the effects of dark matter particle phenomenology on structure formation is usually done using cosmological simulations. However these simulations can only access a finite range of objects due to their limited resolution. An alternative to simulations, originally pioneered by \citet[hereafter PS]{1974ApJ...187..425P} and later extended by use of Excursion Set Theory \citep{1991ApJ...379..440B} and consideration of ellipsoidal collapse \citep{2001MNRAS.323....1S}, is used to calculate `analytical' mass functions. This is done using linear theory to evolve the density field, and assuming a simple model for gravitational collapse in which regions denser than some density threshold collapse to form virialised objects. Using the PS formalism is attractive as it allows us to look at all scales and redshifts simultaneously, while we can easily separate the contribution from haloes of different masses to quantities such as the mean scattering rate for SIDM particles through cosmic time.

This work follows a similar procedure to \citet{2009JCAP...10..009C}, who estimated the DM annihilation rate through cosmic time. The rate of interactions in a DM halo can be calculated given a particle model and the density profile of the halo. Then with a mass function (from PS theory or equivalent) it is possible to work out the total rate of scattering in the Universe. For the simplest model of particle annihilation the DM cross-section, $\sigma$, multiplied by the relative velocity of particles, $v$, is constant. As the rate of interactions is proportional to $\left< \sigma v \right>$ this simplifies the calculation relative to a case where $\sigma$ has some other velocity dependence. In this work we use DM models that have interaction cross-sections that differ from $\sigma \propto 1/v$, first using the simplest model for particle scattering in which $\sigma$ is a constant.

Our study is aimed at estimating the rate of scattering in DM haloes of different masses through cosmic time. The high redshift Universe is very dense, and were it to turn out that the scattering rate was therefore high, the survival of the first seeds of structure formation could provide a useful constraint on the self-interaction cross-section of DM. This work should also be helpful when assessing the importance of resolution in $N$-body simulations of SIDM, because they can only resolve objects above a certain mass. While only the resolved objects from simulations are usually of interest, objects build up in a hierarchical fashion, such that resolved objects at some epoch, are made from the merging of smaller (potentially unresolved) objects from an earlier time. It is therefore important to assess whether these small objects that merged should have been affected by DM self interactions.

The paper is organised as follows. In Section \ref{sect:Cosmic_Scatter_rate} we discuss the calculation of the DM interaction rate through cosmic time for a velocity-independent scattering cross-section, while in Section \ref{sect:varying_stuff} we show the effects of changing the models and parameters that went into our original calculation. In Section \ref{sect:vdSIDM} we perform the same calculation with velocity-dependent cross-sections, focussing in particular on two models recently simulated by \citet{2013MNRAS.430.1722V}. Finally, we give our conclusions in Section \ref{sect:conclusions}. Throughout the paper we assume a Planck 2013 cosmology \citep{2014A&A...571A..16P} unless stated otherwise, and also assume that self-interactions do not effect large scale structure formation.

\section{Interaction rate over cosmic time}
\label{sect:Cosmic_Scatter_rate}

In this section we first discuss the number density of DM haloes of different masses and how this evolves with redshift. By then looking at the scattering rate of DM particles in the haloes that exist at a particular redshift we can calculate the rate of DM particle scattering at that epoch. This calculation assumes that DM scattering is only between particles within the same DM halo, and neglects the fact that scattering rates would be enhanced during the merging of DM haloes, when the relative velocities between particles can be larger. As haloes only spend a small fraction of time undergoing major mergers, the contribution of mergers to the integrated number of scatterings should not be too significant.

\subsection{Mass function of collapsed structures}
\label{sect:Mass_Func}

We initially calculate the number of structures of a given mass using Press-Schechter (PS) theory, considering alternative formulations in Section \ref{sect:PS_vs_ST}. The primordial fluctuations $\delta = (\rho - \mrho) / \mrho$ in the Universe's matter density field $\rho$, are evolved using linear theory.  The spherical collapse model \citep[e.g.][]{1993MNRAS.262..627L} shows that volumes of radius $R$ in which the mean overdensity $\delta_R$ exceeds a critical threshold $\delta_R > \delta_{c}=1.686$ will collapse under their own gravity. We assume gravitational collapse to be immediate leading to a virialised halo with mass $M=\frac{4}{3} \pi R^{3} \mrho$.

To find these volumes, consider smoothing the density distribution on a scale $R$. Assuming the density fluctuations form a Gaussian random field, the fraction of the Universe in regions with an overdensity greater than $\delta_{c}$ is
\begin{equation}
\label{F_delta_R}
F(\delta_{R}>\delta_{c}) = \int_{\delta_{c}}^{\infty} \frac{1}{\sqrt{2 \pi \sigma_{R}^{2}}} \exp \left( -\frac{\delta_{R}^{2}}{2\sigma_{R}^{2}}\right) \, \rmd \delta_{R}.
\end{equation} 
This depends only on $\sigma_{R}^{2}$, the variance of $\delta_{R}$ on this scale.
Because $\delta_{R}$ has zero mean, 
\begin{equation}
\label{rms_M_k}
\sigma_{R}^{2} = \left<  \delta_{R}^{2} \right> = D^{2}(z) \int  k^{2} P(k) \widetilde{W}_{R}^{2}(k) \, \rmd k,
\end{equation}
where the linear growth factor $D(z)$ governs the amplitude of perturbations at redshift $z$, and $\widetilde{W}_{R}(k)$ is the Fourier Transform of a real-space spherical top hat filter of radius $R$. 

The power spectrum $P(k)$ is obtained by multiplying the power spectrum of fluctuations generated by inflation by the Transfer Function $T(k)$, which accounts for the different behaviour of fluctuations that are smaller than and larger than the horizon during the radiation and then matter dominated eras. For simplicity we use the \citet{1998ApJ...496..605E} zero-baryon CDM model in which  
\begin{align}
\begin{split}
T(q) &= \frac{L_0}{L_0 + C_0 \, q^2}, \\
L_0(q) &= \ln(2e+1.8q), \\
C_0(q) &= 14.2 + \frac{731}{1+62.5q},
\label{T_k}
\end{split}
\end{align}
and $q$ is related to $k$ by
\begin{equation}
\label{q_k}
q = \frac{k}{\Omega_m \, h^2 \, \mathrm{Mpc}^{-1}} ~\Theta_{2.7}^{2}\, ,
\end{equation}
where $T_\mathrm{CMB} = 2.7 \, \Theta_{2.7} \, \mathrm{K}$. We look at the effect of changes to the high-$k$ power spectrum by integrating the mass function down to different minimum masses, as explained in Section \ref{sect:DM_cosmic_interaction_rate}.

\begin{figure}
        \centering
        \includegraphics[width=\columnwidth]{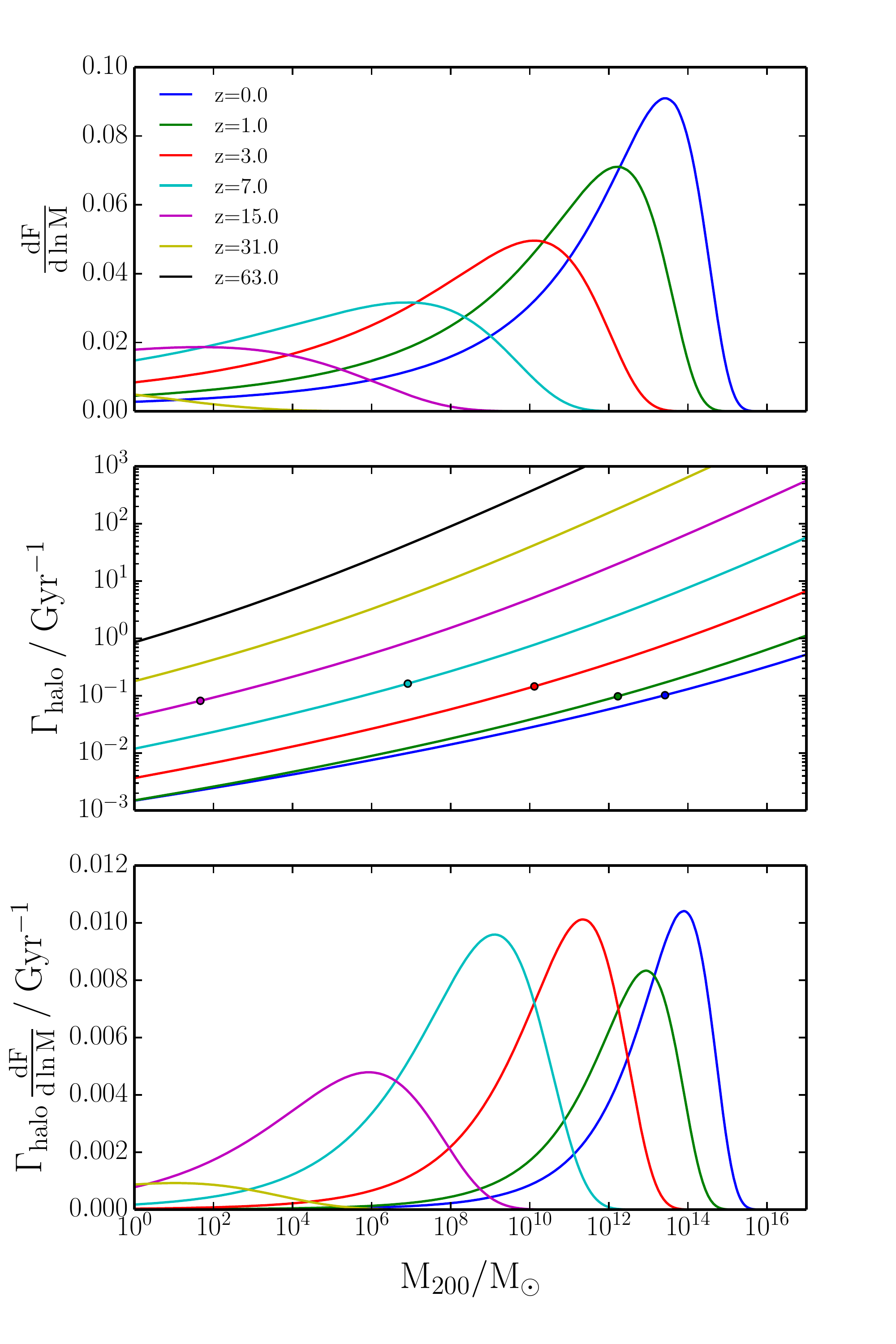}
	\caption{
	{\it Top panel}: The Multiplicity Function,
	which shows how mass in the Universe is split between objects of different mass, as described by Press-Schechter theory. Different coloured lines show different redshifts.
	{\it Middle panel}: The interaction rate per particle as a function of halo mass, assuming NFW density profiles and the \citet{2008MNRAS.390L..64D} concentration-mass relation, with a velocity-independent cross-section of $1\, \mathrm{cm^2 \, g^{-1}}$. Circles highlight the mass at which the Multiplicity Function peaks, illustrating a relatively constant interaction rate per unit mass in the Universe's most typical haloes.
	{\it Bottom panel}: The product of curves in the two upper panels, illustrating the relative contribution of haloes in different logarithmic mass bins to the total interaction rate per particle.
	In this scenario, the main location for scatterings gradually transitions to more and more massive structures.}
	\label{fig:Duffy_3panel}
\end{figure}

PS theory then interprets the fraction of the Universe's volume for which $\delta_R > \delta_c$ as the fraction of the Universe's mass that has collapsed to form objects with mass $M\geq\frac{4}{3} \pi R^{3} \left< \rho \right>$.
In this transition from smoothing over volumes to mass scales, it is also convenient to eliminate time dependence from the rms density fluctuations, i.e. we define the rms mass fluctuations on scale $M$ as $\sigma_{M}\equiv\sigma_{R}(z)/D(z)$, such that $D(z\!=\!0) = 1$.
Thus the fraction of the mass in the Universe in collapsed objects with mass greater than $M$, at redshift $z$, is
\begin{equation}
\label{F_M}
F(M,z) = \int_{\delta_{c}/\sigma_{M}D(z)}^{\infty} \frac{1}{\sqrt{2 \pi }} \exp \left( -\frac{\xi^{2}}{2}\right) \, \rmd\xi,
\end{equation}
where $\xi = \delta_{M} / \sigma_{M} D(z)$. 
This depends only on the rms density fluctuations (in the lower limit of integration) and the linear growth factor. 
Differentiating it with respect to mass yields the \emph{multiplicity function}
\begin{equation}
\label{F_logM}
\frac{\rmd F}{\rmd \,\ln M}(z) = \sqrt{\frac{2}{\pi}} \left| \frac{\rmd \, \ln \sigma_M}{\rmd \, \ln M} \right| \nu \exp \left( -\frac{\nu^{2}}{2} \right),
\end{equation}
where we have introduced $\nu = \delta_{c}/\sigma_{M}D(z)$ and multiplied by a factor of two to account for mass that is initially in under-dense regions.\footnote{Consider what happens when we take $M \to 0$ in Equation \ref{F_M}. On small scales the rms fluctuations are very large, and the lower limit in the integration tends to zero. This implies $F(0,z) = \frac{1}{2}$, and only half of the mass in the Universe is in collapsed objects. On small enough scales the density field is always non-linear, and so we would expect all mass in the Universe to be in collapsed objects if we take $M \to 0$. The missing half of the Universe corresponds to regions that are below the collapse threshold when smoothed on a scale $M$, but would be above the collapse threshold if smoothed on some larger scale. For more information, see the discussion of the `cloud-in-cloud' problem in \citet{1991ApJ...379..440B}.}
This describes how the mass in the Universe is divided amongst objects of different mass and is plotted in the top panel of Fig. \ref{fig:Duffy_3panel}.

\subsection{Interaction rates in collapsed structures}
\label{sect:Gamma_halo}

The scattering rate of an individual dark matter (DM) particle $i$, with velocity $\boldsymbol{v_{i}}$, is
\begin{equation}
\Gamma_{i} = \int f(\boldsymbol{v'}) \, \rho \, \frac{\sigma}{m} \, |\boldsymbol{v_{i}} - \boldsymbol{v'}| \, \rmd^{3} \boldsymbol{v'},
\label{Gamma_i}
\end{equation}
where $f$ is the velocity distribution function,\footnote{Here $f$ is normalised such that $\int f(\boldsymbol{v_{1}}) \, d^{3}\boldsymbol{v_{1}} = 1$.} $\rho$ the local density, and $(\sigma/m)$ the cross-section for DM-DM scattering (which could depend on $|\boldsymbol{v_{i}} - \boldsymbol{v'}| \equiv v_\mathrm{pair}$) divided by the DM particle mass. 
Integrating over the velocity distribution function gives the scattering rate for a particle at position $\boldsymbol{r}$,
\begin{equation}
\Gamma_{i}(\boldsymbol{r}) = \frac{\langle \sigma \, v_\mathrm{pair}\rangle(\boldsymbol{r})  \rho(\boldsymbol{r}) }{m}.  \label{Gamma_i_r}
\end{equation}

For a halo of mass $M$ containing $N$ particles, the mean scattering rate per particle is 
\begin{equation}
\Gamma_\mathrm{halo}(M)=\frac{1}{N} \sum_{i=1}^N \Gamma_{i}.
\end{equation}
Integrating over radius $r$ gives
\begin{equation}
\Gamma_\mathrm{halo}(M)=\frac{1}{M} \int_{0}^{\infty} 4 \pi r^2 \rho(r) \Gamma_{i}(r) ~ \rmd r
\end{equation}
\begin{equation}
\hspace{14mm}=\frac{1}{M} \int_{0}^{\infty} 4 \pi r^2 \rho^2(r) \frac{\langle \sigma\, v_\mathrm{pair}\rangle(r)}{m}  \, \rmd r.
\label{Gamma_halo}
\end{equation}

We assume that the collapsed haloes from PS theory have spherically symmetric \citet[][hereafter NFW]{1997ApJ...490..493N} density profiles,
\begin{equation}
\frac{\rho(r)}{\rho_\mathrm{crit}} = \frac{\delta_\mathrm{NFW}}{(r/r_s)(1+r/r_s)^2},
\end{equation}
where $r_s$ is a scale radius, $\delta_\mathrm{NFW}$ a dimensionless characteristic density, and $\rho_\mathrm{crit}=3H^2/8\pi G$ is the critical density. 
We assume that the mass of a halo fills a spherical region of radius $r_{200}$, within which the mean density is $200\, \rho_\mathrm{crit}$ and the total mass is $M_{200}$.
Outside this region we assume the density to be zero. For brevity we will also refer to $r_{200}$ as $r_{\mathrm{v}}$ and $M_{200}$ as $M$. The concentration parameter is defined as $c\equiv r_{200}/r_s$ and can be related to the characteristic density by
\begin{equation}
\delta_\mathrm{NFW} = \frac{200}{3}\frac{c^3}{\ln (1+c)-c/(1+c)}.
\end{equation}

Note that the NFW profile is obtained from non-interacting dark matter simulations. 
Dark matter scattering reduces the density in the centre of DM haloes, producing a constant density core \citep{2000ApJ...534L.143B,Yoshida:2000gn,2000PhRvL..84.3760S,2002ApJ...581..777C,Rocha:2013bo,2013MNRAS.431L..20Z}. Assuming an NFW profile, the average radius at which interactions take place (assuming an isotropic velocity dispersion) is $0.32 \, r_s$ independent of halo concentration. This is similar to the radius for which the radial density profiles seen in the simulations of \citet{Rocha:2013bo} first drop below the NFW prediction. These simulations used the maximum allowed velocity-independent cross-section, and so cores in other models would likely be smaller. Also, while the density in the centres of haloes decreases, DM scattering increases the velocity dispersion in halo centres, which should cancel some of the effect. Nevertheless we acknowledge that these DM interactions are moderately self-regulating because they form cores that will tend to decrease the interaction rate, but proceed assuming an NFW profile for the DM density. If we relax this assumption then the scattering rates calculated would be lower, but a full treatment of the effect that scattering has on the phase space distribution of haloes, and so the subsequent scattering rates, requires full N-body simulations that are beyond the scope of this paper.

To calculate the mean pairwise velocity of particles, we integrate over the velocity distribution functions of particle pairs.
Assuming that their velocities are isotropic
and follow a Maxwell-Boltzmann distribution\footnote{High resolution simulations of CDM report departures from Gaussianity for the distribution of velocity components along the principal axes of the velocity dispersion tensor \citep{2009MNRAS.395..797V}, but this approximation is sufficient for our work.} with one dimensional velocity dispersion $\sigma_{\mathrm{1D}}$, this gives
$\langle v_\mathrm{pair}\rangle=(4/\sqrt{\pi})\, \sigma_{\mathrm{1D}}$.
For an NFW halo, the velocity dispersion of particles is \citep{2001MNRAS.321..155L}
\begin{equation}
\label{eq:v_disp}
\begin{split}
\sigma_{\mathrm{1D}}^{2}(s,c) 	&= \frac{1}{2} c^2 g(c) s (1+cs)^2 \frac{G M}{r_{\mathrm{v}}} \left[ \pi^2 - \ln(cs) - \frac{1}{cs} \right. \\
				&- \frac{1}{(1+cs)^2} - \frac{6}{1+cs} + \left( 1 + \frac{1}{c^2s^2} - \frac{4}{cs} - \frac{2}{1+cs} \right) \\
				&\times \left.\vphantom{\frac12} \ln(1+cs) + 3 \ln^2(1+cs) + 6\mathrm{Li}_{2}(-cs) \right],
\end{split}
\end{equation}
where $s\equiv r/r_{200}$, $g(c)\equiv[\ln (1+c) - c/(1+c)]^{-1}$,
and $\mathrm{Li}_{2}(x)$ is the dilogarithm (commonly referred to as Spence's function), defined by 
\begin{equation}
\mathrm{Li}_{2}(x) = \int_x^0 \frac{\ln (1-u)}{u} \, \rmd u.
\end{equation}

Returning to Equation~\eqref{Gamma_halo}, and changing integration variable from $r$ to $s$, we find
\begin{equation}
\label{Gamma_halo_s}
\Gamma_\mathrm{halo}(M,\rv,c)= 16 \sqrt{\pi} \frac{\rv^{3}}{M} \frac{\sigma}{m}  \int_{0}^{1} s^2 \rho^2(s,c)\, \sigma_{\mathrm{1D}}(s,c)\, \rmd s,
\end{equation}
where we have now assumed that the DM-DM cross-section is velocity-independent (this restriction is relaxed in Section \ref{sect:vdSIDM}). 
Both $\rho(s,c)$ and $\sigma_{\mathrm{1D}}(s,c)$ depend on the virial mass and radius of a halo, and can be written as dimensionless functions of $s$ and $c$ multiplied by the dimensional quantities $M / \rv^3$ and $\sqrt{G M / \rv}$ respectively. We can then see that $\Gamma_\mathrm{halo}$ will be a function of the halo concentration scaled by power-laws in $M$ and $\rv$. Specifically, at fixed cross-section and halo concentration, $\Gamma_\mathrm{halo} \propto M^{3/2} \, \rv^{-7/2}$.

At a particular cosmic time, $M \equiv M_{200}$ and $\rv \equiv r_{200}$ are not independent, because  $M_{200} / r_{200}^{3} \propto \rho_\mathrm{crit}(z)$ by definition. Using this, we find $\Gamma_\mathrm{halo} \propto M^{1/3} \, \rho_\mathrm{crit}^{7/6}$, such that
\begin{equation}
\begin{split}
\label{Gamma_halo_scaling}
\Gamma_\mathrm{halo}(M,z,(\sigma/m),c) 	&= \Gamma_\mathrm{halo}(M_0,z_0,(\sigma/m)_0,c) \left(\frac{M}{M_0}\right)^{1/3} \\
							&\times  \left(\frac{\rho_\mathrm{crit}(z)}{\rho_\mathrm{crit}(z_0)}\right)^{7/6} \left(\frac{(\sigma/m)}{(\sigma/m)_0}\right).
\end{split}
\end{equation}
We calculate $\Gamma_\mathrm{halo}(M_0,z_0,(\sigma/m)_0,c)$ with $M_0=10^{10}\, \rmm_\odot$, $z_0=0$ and $(\sigma/m)_0 = 1\, \mathrm{cm^2 \, g^{-1}}$, by numerically integrating Equation~\eqref{Gamma_halo_s}. We can then calculate $\Gamma_\mathrm{halo}$ for haloes with different masses and at different redshifts using Equation \eqref{Gamma_halo_scaling}.

At fixed mass, redshift and cross-section, $\Gamma_\mathrm{halo}$ is found to increase significantly with increasing halo concentration. The logarithmic slope of the $\Gamma_\mathrm{halo}(c)$ relation is $\sim 1.7$ for $c=5$, and $\sim 2.5$ for $c=30$, with $\Gamma_\mathrm{halo} \propto c^2$ for concentrations around 10. As halo concentrations generally decrease with increasing halo mass, the mass dependence of $\Gamma_\mathrm{halo}$ is suppressed below the $\Gamma_\mathrm{halo} \propto M^{1/3}$ seen in Equation \eqref{Gamma_halo_scaling}. The overall form of $\Gamma_\mathrm{halo} (M,z)$ depends upon the concentration-mass-redshift relation.
Following \citet[hereafter D08]{2008MNRAS.390L..64D}, we shall initially assume 
\be
c(M,z)=5.72 \left( \frac{M}{10^{14} \, h^{-1} \rmm_\odot} \right)^{-0.081} (1+z)^{-0.71}.
\ee
Using this $c(M,z)$ relation we show $\Gamma_\mathrm{halo}(M,z)$ in the middle panel of Fig.~\ref{fig:Duffy_3panel}. $\Gamma_\mathrm{halo}$ increases rapidly with increasing redshift at fixed mass, and increases with mass at fixed redshift. As objects grow in mass through cosmic time, the scattering rate in typical haloes at each redshift evolves slowly. Note that several more recent works show that this simple power law dependence of $c(M)$ should flatten at low masses, as is discussed in Section \ref{sect:c_Mz}.

\subsection{DM's cosmic scattering rate}
\label{sect:DM_cosmic_interaction_rate}

\begin{figure}
        \centering
        \includegraphics[width=\columnwidth]{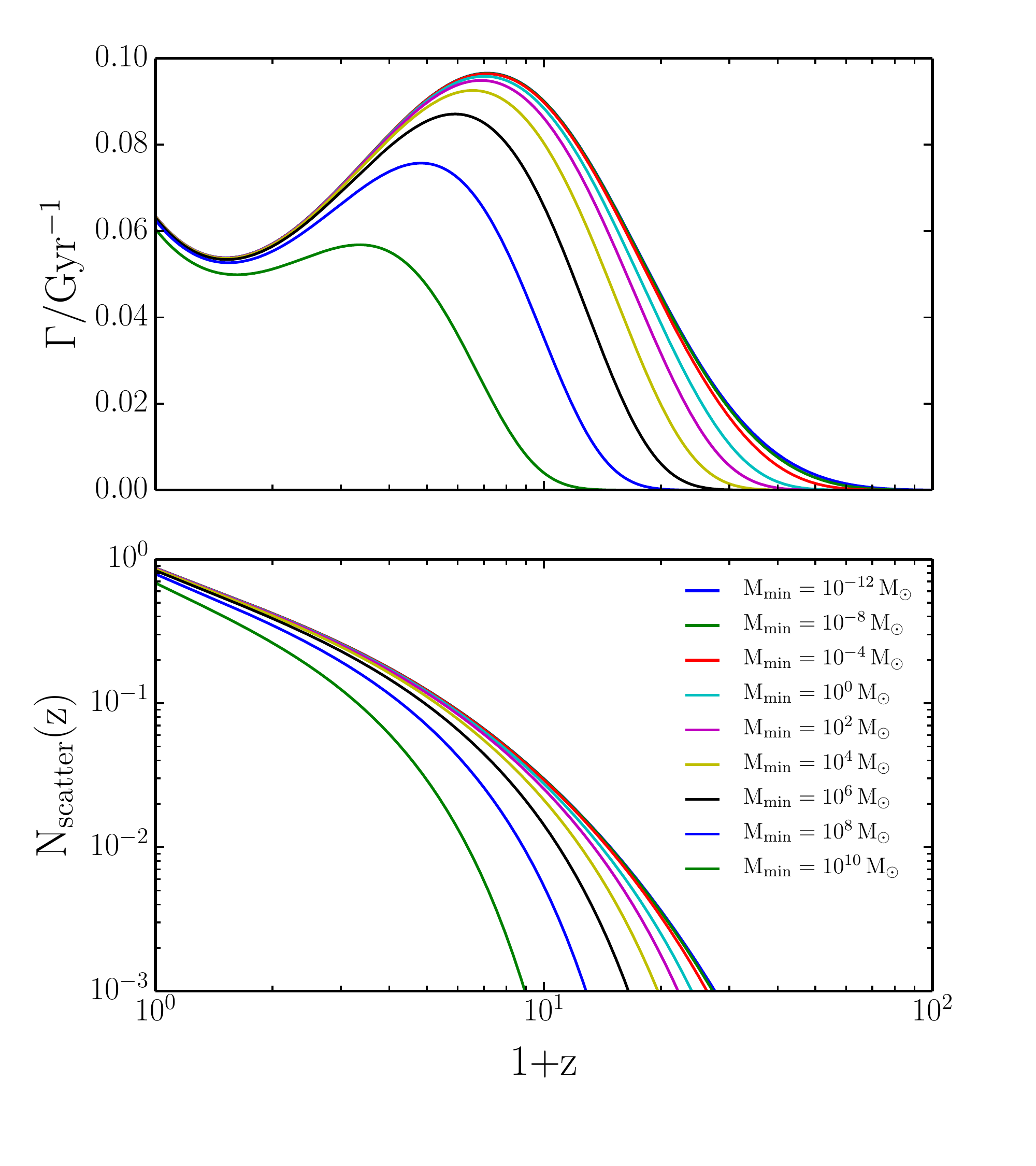}
	\caption{{\it Top panel}: The mean scattering rate of particles in the Universe calculated from Press-Schechter theory, assuming the NFW density profile, the D08 concentration-mass relation and $\sigma / m =1\, \mathrm{cm^2 \, g^{-1}}$. The different lines count only scatterings in haloes more massive than $10^{10} \rmm_{\odot}$ (bottom line) to $10^{-12} \rmm_{\odot}$ (top line). The scattering rate varies by less than a factor of two from $z \approx 6$ onwards.
	{\it Bottom panel}: The mean cumulative number of interactions that particles have undergone as a function of redshift. The different lines again include only those interactions in haloes more massive than a given threshold. With a velocity-independent cross-section, most scattering is at late redshifts where there is more time. This results in most scattering being in high-mass haloes, so that $N_\mathrm{scatter}(z=0)$ varies by less than 25\% between $M_\mathrm{min}=10^{-12} \rmm_{\odot}$ and $M_\mathrm{min}=10^{10} \rmm_{\odot}$.}
	\label{fig:cosmic_scatter_rate}
\end{figure}

Multiplying the multiplicity function from Section \ref{sect:Mass_Func} by the interaction rate in individual haloes from Section \ref{sect:Gamma_halo} gives the contribution of haloes of different mass to the total rate of particle scattering in the Universe (see bottom panel of Fig.~\ref{fig:Duffy_3panel}).
Integrating this quantity over all halo masses at different redshifts yields the mean scattering rate of all particles at that redshift, $\Gamma(z)$, which we refer to as the `Cosmic Scattering Rate'. This is plotted in Fig.~\ref{fig:cosmic_scatter_rate}, where it can be seen that after a gradual rise from the early universe to $z \approx 6$, $\Gamma(z)$ is constant to within a factor of two to the present day.

For this analysis, we assume that haloes form down to masses of $10^{-12} \, \rmm_\odot$. 
In the real Universe, self-interacting dark matter creates a small-scale cut-off in the power spectrum due to collisional damping. For DM composed of weakly interacting massive particles (e.g.~neutralinos), the minimum mass of collapsed objects is $\sim 10^{-6}\,\rmm_{\odot}$  \citep{2001PhRvD..64h3507H}. If DM were axions then this minimum mass would be $\sim 10^{-12}\,\rmm_{\odot}$ \citep{1996ApJ...460L..25K}.
For the general class of self-interacting dark matter models that lead to astrophysically interesting scattering rates in the late-time Universe, collisions in the early Universe suppress power on larger scales, or even introduce acoustic oscillations in the dark matter-dark radiation system \citep{2014PhRvD..90d3524B}. There is a rich possible phenomenology affecting the high-$k$ power spectrum, which is highly model-dependent.

We investigate the approximate effect of a cutoff in the power spectrum by integrating $\Gamma_\mathrm{halo}(\rmd F/\rmd \ln M)$ down to different minimum masses, $M_\mathrm{min}$, shown as the extra lines in Fig.~\ref{fig:cosmic_scatter_rate}. Furthermore, in numerical simulations, only haloes above a given mass scale are resolved, and only the DM interactions above those scales can be tracked. We therefore include lines with large $M_\mathrm{min}$ in Fig.~\ref{fig:cosmic_scatter_rate}, to act as predictions for the expected scattering rate in cosmological simulations. Note that the results as $M_\mathrm{min} \to 0$ converge particularly slowly for the D08 concentration-mass relation, due to the high concentration of very small haloes. Nevertheless, these results are less sensitive to changing $M_\mathrm{min}$ than those for a simple annihilation channel where $\sigma v_\mathrm{pair}$ is constant \citep{2014MNRAS.439.2728M} and low mass haloes make a dominant contribution to the total scattering rate.

In addition to the rate of cosmic scattering, an interesting quantity is the mean cumulative number of interactions that particles have undergone.  As each scattering event is a two-body interaction, this is twice the number of interactions per particle. We call this quantity $N_\mathrm{scatter}$ and plot it as a function of redshift in the bottom panel of Fig.~\ref{fig:cosmic_scatter_rate}. While the cosmic scattering rate is markedly different at intermediate and high redshifts when using different minimum masses, the values of $N_\mathrm{scatter}(z=0)$ are more robust. For $(\sigma / m)=1\, \mathrm{cm^2 \, g^{-1}}$, $N_\mathrm{scatter}(z=0)$ is $0.87$ with $M_\mathrm{min}=10^{-12}\, \rmm_{\odot}$ and $0.68$ with $M_\mathrm{min}=10^{10}\, \rmm_{\odot}$. 

A noticeable feature of $\Gamma(z)$ in the upper panel of Fig. \ref{fig:cosmic_scatter_rate} is the upturn after $z \approx 1$. This is not present when using more recent $c(M,z)$ relations with more complex redshift dependences than the simple $(1+z)^{-0.71}$ in the D08 relation. This upturn is not physical, and arises because the concentration is defined in terms of $r_{200}$ which in turn depends on $\rho_\mathrm{crit}$. When the Universe is matter-dominated $\rho_\mathrm{crit} \propto (1+z)^3$, such that at fixed halo mass $r_{200} \propto (1+z)^{-1}$. At late times, when there is a significant dark energy contribution to the Universe, the evolution of $\rho_\mathrm{crit}$ slows and is no longer given by a simple power law in $(1+z)$. This affects the $r_{200}$ of haloes, and hence halo concentrations, such that a simple power-law cannot accurately capture $c(M,z)$.

\section{Sensitivity to Astrophysical Assumptions}
\label{sect:varying_stuff}

In the previous section we considered the redshift dependence of DM scattering rates and showed that with a velocity-independent cross-section, the mean rate of particle scattering in the Universe initially grows and then starts to decrease after $z \approx 6$, dropping by less than a factor of two to the present day. In this section we explore the sensitivity of this result to the assumptions of the model.

\subsection{Concentration-mass-redshift relations}
\label{sect:c_Mz}

The concentration-mass-redshift relation, $c(M,z)$, of D08 is attractive for its simplicity and because over a small range of redshifts and halo masses, concentrations can be well fit by simple power laws in $M$ and $(1+z)$. However, numerical studies that have resolved structures over a wide range of halo masses have found that concentrations are not well fitted by simple power laws. Examining the results of the Millennium Simulation \citep{2005Natur.435..629S} from $z=3$ to $z=0$ it is clear that the form of $c(M,z)$ is not separable, with the mass dependence evolving with redshift \citep{2008MNRAS.387..536G}. This evolution takes the form of a flattening of the $c(M)$ relation at increasing redshift, such that concentrations of very massive galaxy cluster haloes evolve only weakly with redshift while the concentrations of smaller haloes decrease rapidly with increasing redshift.  

The $c(M,z)$ relation is found to be remarkably complex, particularly when considering the dependence on cosmological parameters. \citet[hereafter P12]{2012MNRAS.423.3018P} find that this complex relationship is a result of the `wrong' physical quantities, $M$ and $z$, being used. Analogous to studies of the halo mass function, in which a much simpler fitting formula is possible when one considers the mass function as a function of $\ln \sigma_M^{-1}$ rather than a function of $M$ \citep{2001MNRAS.321..372J}, the $c(\ln \sigma_M)$ relationship is found to be simpler than $c(M)$.

The behaviour of this relationship can be explained by models in which the concentration of a halo is related to its accretion history \citep{2002ApJ...568...52W,2003MNRAS.339...12Z}. \citet[hereafter L14]{2014MNRAS.441..378L} found that if the mass of a halo, $M(z)$, was plotted against the critical density, $\rho_{\mathrm{crit}}(z)$, then the relationship $M(\rho_{\mathrm{crit}})$ was well fit by an NFW profile, with associated concentration $c_{\mathrm{MAH}}$. They also found a simple relation between $c_{\mathrm{MAH}}$ and the concentration of the halo, allowing the concentration-mass relation to be predicted from the mass-accretion history of haloes. The statistics of the mass-accretion of DM haloes can be found from simulations, or calculated using the conditional probabilities\footnote{The conditional probability that the material making up an object of mass $M_1$ at redshift $z_1$ is in an object of mass $M_0$ at redshift $z_0$.} found in extensions of PS theory \citep{1991ApJ...379..440B, 1991MNRAS.248..332B, 1993MNRAS.262..627L, 1993MNRAS.261..921K}. 

\begin{figure}
        \centering
        \includegraphics[width=\columnwidth]{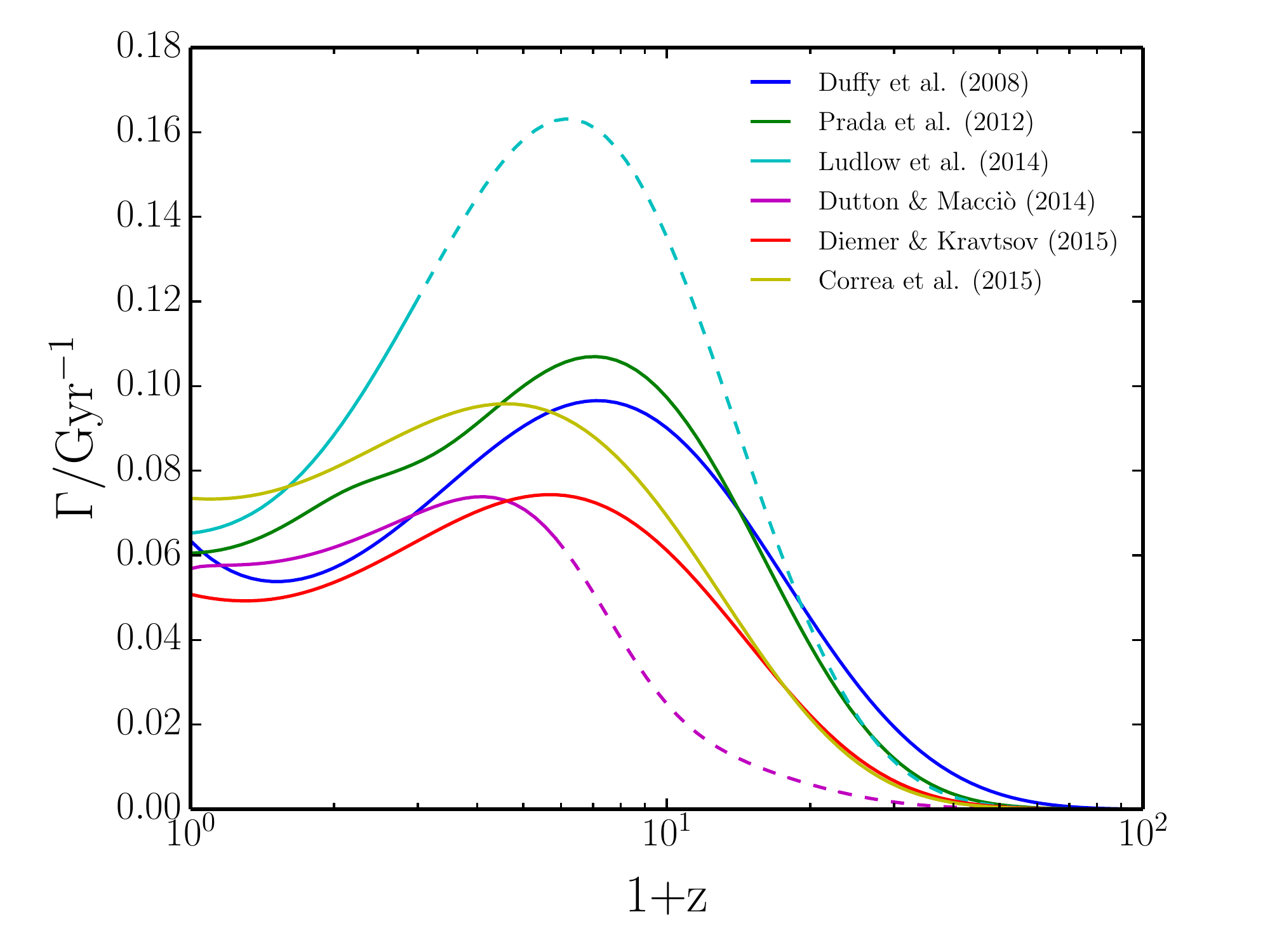}
	\caption{ The cosmic scattering rate calculated using the concentration-mass-redshift relations of
	\citet{2008MNRAS.390L..64D},
	\citet{2012MNRAS.423.3018P},
	\citet{2014MNRAS.441..378L},
	\citet{2014MNRAS.441.3359D},
	\citet{2015ApJ...799..108D}, and 
	\citet{2015MNRAS.452.1217C}.
	These were calculated assuming a Planck 2013 cosmology \citep{2014A&A...571A..16P}, a PS mass-function, and $\sigma / m =1\, \mathrm{cm^2 \, g^{-1}}$, counting the contribution from all haloes more massive than $10^{-12} \, \rmm_\odot$. Lines become dashed for redshifts where authors state their relationships may not be valid.}
	\label{fig:cM_cosmic_scattering}
\end{figure}

Different methods for measuring $c(M,z)$, either from simulations or analytical calculations, give similar results around the peak of the multiplicity function ($M \approx M^*$), but differ significantly at high and low masses. While the cosmic scattering rate is dominated by haloes around $M^*(z)$, the scattering rate in haloes is highly sensitive to the halo concentration, and so even small differences between $c(M,z)$ relations can lead to significant changes in $\Gamma(z)$. In Fig.~\ref{fig:cM_cosmic_scattering} we show $\Gamma(z)$ calculated as in Fig.~\ref{fig:cosmic_scatter_rate} but for a variety of $c(M,z)$ relations.

Noticeable in Fig. \ref{fig:cM_cosmic_scattering} is that using $c(M,z)$ from L14 gives a scattering rate at intermediate redshifts a factor of two above that found using other $c(M,z)$ relations. The L14 analytical model was calculated for relaxed haloes, which are generally dynamically older, making them more concentrated than unrelaxed haloes of a similar mass. The cuts made to remove unrelaxed haloes are one of the two main reasons why $c(M,z)$ relations from simulations disagree with each other, the other being the way in which $c$ is calculated from a mass distribution. For example, \citet{2012MNRAS.423.3018P} calculate $c$ from the ratio $V_{\mathrm{max}}/V_{200}$, where $V_{\mathrm{max}}$ and $V_{200}$ are the maximum circular velocity and the circular velocity at $r_{200}$ respectively, while \citet{2015ApJ...799..108D} find $c$ by directly fitting the radial density with an NFW profile.

\subsection{Mass Function Prescription}
\label{sect:PS_vs_ST}

It is known that the PS formula does not provide an exact fit to the mass function from simulations. Specifically, it underestimates the number of rare objects in the `high-mass tail', with an overestimate of the amount of mass around the peak of the multiplicity function \citep[see e.g.][]{2001MNRAS.321..372J}. A better fit to the mass function from simulations was achieved by \citet[hereafter ST]{1999MNRAS.308..119S}, who found that compared to PS, Equation \eqref{F_logM} becomes:
\begin{equation}
\label{ST_F_logM}
\frac{\rmd F}{\rmd \, \ln M} = A\sqrt{\frac{2a}{\pi}} \left| \frac{\rmd \ln \sigma_M}{\rmd \ln M} \right| \left[ 1 + \left( a \nu^{2}\right)^{-p} \right] \nu \exp \left( -\frac{a \nu^{2}}{2} \right),
\end{equation}
with $A=0.3222$, $a=0.707$ and $p=0.3$. We note that our definition of $\nu$ is different from that in the ST paper, with $\nu_{ST} = \nu^{2}$. The original PS formula can also be described by Equation \eqref{ST_F_logM} with $A=0.5$, $a=1$ and $p=0$.

The ST mass function increases the number density of the most massive objects compared to the PS mass function, providing a better fit to simulations \citep[see e.g.][]{2007MNRAS.374....2R}. While these differences can be extremely important for some studies (e.g. counting the number density of massive clusters) we find that the different mass functions do not have a large effect on our results. This is because the scattering rate per unit mass in DM haloes increases only gently with increasing halo mass, as can be seen in the middle panel of Fig. \ref{fig:Duffy_3panel}. The shape of $\Gamma(z)$ is similar when either a PS or ST mass function is used, although the normalisation is slightly lower for the latter. By redshift zero there are $\sim 20 \%$ fewer DM interactions with an ST mass function.

\subsection{Varying Cosmological Parameters}

Similar to changing the formalism used to calculate the multiplicity function, small changes to the Cosmological Parameters leave the cosmic scattering rate relatively unchanged because of the weak mass dependence of $\Gamma_\mathrm{halo}(M)$. Using $c(M,z)$ from D08, we find that changing cosmological parameters from Planck 2013 to WMAP9 decreases the mean number of interactions per particle by redshift zero, $N_\mathrm{scatter}(z=0)$, by 12\%. This is driven by Planck's larger value for $\Omega_\mathrm{m}$, resulting in larger critical densities at early times. Using earlier WMAP results leads to similar changes, except for WMAP3 for which the anomalously low $\Omega_\mathrm{m}$ and $\sigma_8$ lead to a 33\% reduction in $N_\mathrm{scatter}(z=0)$.

The concentration-mass-redshift relation also depends on cosmological parameters, which is made explicitly clear by relations that relate $c$ to $\sigma_M$ rather than $M$ directly \citep[e.g.][]{2012MNRAS.423.3018P,2015ApJ...799..108D}. This cosmology dependence of $c(M,z)$ makes little difference when moving from Planck 2013 to WMAP9, but further reduces the scattering rate for a WMAP3 cosmology such that $N_\mathrm{scatter}(z=0)$ is 40\% lower than with a Planck 2013 cosmology, using $c(M,z)$ from P12. This increased difference, beyond that seen for a cosmology independent $c(M,z)$, can be understood by noting that haloes of a particular mass form later with smaller $\sigma_8$, and are therefore less concentrated.

\subsection{Scatter in the Concentration-Mass Relation}

So far we have assumed that given the mass of a halo we know its concentration through the concentration-mass relation. In practice this relation has some scatter around it, which will impact on the mean scattering rate of haloes. From Equation \eqref{Gamma_halo_scaling} the concentration dependence of the scattering rate in haloes is described by $\Gamma_\mathrm{halo}(M_0,z_0,(\sigma/m)_0,c)$. This is non-linear in $c$, such that even symmetric scatter in $c$ at fixed mass will alter the mean scattering rate in haloes of that mass. 

To discuss how $\Gamma_\mathrm{halo}$ is affected by scatter in $c$, it will be useful to introduce $c_0$, the value of $c$ implied by the $c(M,z)$ relation. \citet{2004A&A...416..853D} find that for haloes of fixed mass and redshift, $\ln c$ is normally distributed. If we assume that $\ln c$ is normally distributed with mean $\ln c_0$ and variance $\sigma_{\ln c}^{2}$, then $c$ follows a log-normal distribution, with probability density function
\begin{equation}
\label{log_Normal}
P(c) = \frac{1}{c \, \sigma_{\ln c} \sqrt{2 \pi}} \exp \left( -\frac{(\ln c - \ln c_0)^2}{2 \, \sigma_{\ln c}^{2}} \right).
\end{equation}

Including a log-normal distribution of concentrations at fixed mass and redshift leads to an increase in $\Gamma$ at all concentrations, related to the long tail of the distribution towards high values, as well as a shift in the expectation value of $c$.\footnote{For the distribution in Equation \eqref{log_Normal}, the expectation value of $c$ is given by $\left< c \right> = \exp \left( \ln c_0 + \sigma_{lc}^{2}/2 \right) > c_0$.} If $c(M,z)$ in D08 was a measure of the mean $c$ for a particular mass of halo, then we would have to make the change $\ln c_0 \to \ln c_0 - \sigma_{\ln c}^{2}/2$ in Equation \eqref{log_Normal} to keep $\left< c \right> = c_0$. However, the $c(M,z)$ relation in D08 was acquired by fitting to the median values of $c$ in each mass bin at each redshift. The median value of $c$ from the probability density function in Equation \eqref{log_Normal} is simply $\exp \left( \ln c_0  \right)=c_0$ as required. \citet{2004A&A...416..853D} found that $\sigma_{\ln c} \approx 0.22$, almost independent of the cosmological model. This corresponds to a standard deviation in $\log_{10} c$ of $0.1$, or a scatter of 0.1 dex. We find that the shape of $\Gamma(z)$ is effectively unchanged by scatter in $c(M,z)$, but that the normalisation increases with increasing scatter. For a 0.1 dex scatter, the normalisation increases above that of the scatter-free case by less than 15\%.

\section{velocity-dependent cross-sections}
\label{sect:vdSIDM}

Having calculated the rate of DM scattering through cosmic time assuming that the cross-section is velocity-independent, we now lift this assumption, and perform the same calculation with velocity-dependent DM-DM cross-sections.

\subsection{Particle model}

For velocity-dependent cross-sections we use the vdSIDMa and vdSIDMb models described in \citet{2013MNRAS.430.1722V}. These are well-motivated by particle physics, and describe the transfer cross-section for elastic scattering mediated by a new gauge boson of mass $m_\phi$. This results in an attractive Yukawa potential with coupling strength $\alpha_c$. These interactions are analogous to the screened Coulomb scattering in a plasma, for which the momentum-transfer cross-section can be approximated by
\begin{equation}
\frac{\sigma_T}{\sigma_T^{\rm max}} \approx 
     \begin{cases}
       \frac{4\pi}{22.7}~\beta^2~{\rm ln}\left(1+\beta^{-1}\right),                    &\beta<0.1\\ \\
       \frac{8\pi}{22.7}~\beta^2~\left(1+1.5\beta^{1.65}\right)^{-1},                   &0.1<\beta<10^3\\ \\
       \frac{\pi}{22.7}~\left({\rm ln}\beta+1-\frac{1}{2}({\rm ln} \, \beta)^{-1} \right)^2,  &\beta>10^3, 
     \end{cases}
\label{eq:cross}
\end{equation}
where $\beta=\pi v_{\rm max}^2/v_\mathrm{pair}^2$ and $\sigma_T^{\rm max}=22.7/m_{\phi}^2$ \citep{2010PhRvL.104o1301F,2011JCAP...05..002F,2011PhRvL.106q1302L}. Here $v_\mathrm{max}$ is the velocity at which $(\sigma_T \, v_\mathrm{pair})$ peaks, with $\sigma_T(v_\mathrm{max}) = \sigma_T^{\rm max}$. We have also introduced the ``momentum-transfer cross-section'', $\sigma_T$, defined as
\begin{equation}
\sigma_T = \int (1 - \cos \theta ) \frac{\rmd \sigma}{\rmd \Omega}(\theta) \, \rmd \Omega
\label{eq:sigma_T1}
\end{equation}
\begin{equation}
\hspace{5mm}= 2 \pi \int_{-1}^{1} (1 - \cos \theta) \frac{\rmd \sigma}{\rmd \Omega}(\theta) \, \rmd \cos \theta,
\label{eq:sigma_T2}
\end{equation}
where $\frac{\rmd \sigma}{\rmd \Omega}$ is the differential cross-section, assumed to be azimuthally symmetric, which describes the probability of particles scattering into a patch of solid angle $\rmd \Omega$.
The transfer cross-section is an effective scattering cross-section that is useful in describing angularly dependent cross-sections (where $\frac{\rmd \sigma}{\rmd \Omega}$ is not constant). For isotropic scattering ($\frac{\rmd \sigma}{\rmd \Omega} = \mathrm{constant}$) the transfer cross-section is simply $\sigma_T = \sigma$, while in general the mean momentum transfer for a scattering process with transfer cross-section $\sigma_T$ is equal to the mean momentum transfer for isotropic scattering with $\sigma = \sigma_T$. Throughout the rest of this paper, when calculating the rate and number of particle scattering events we will use $\sigma_T$ as if it were the cross-section i.e we will calculate an effective rate of particle scatterings that is the rate of isotropic scattering events that would lead to the same rate of momentum transfer.\footnote{In general, particle orbits within a DM halo are approximately isotropic, so there is no preferred direction for particle scattering. In these cases, the momentum transfer cross-section accurately captures the effects of scattering. However, this may not be the case for systems where there is a preferred direction along which particles approach \citep{2014MNRAS.437.2865K}, such as in a merger.}

The velocity-dependent cross-section in Equation \eqref{eq:cross} leads to noticeable changes in $\Gamma_\mathrm{halo}(M)$. The cross-section diverges as the pairwise velocity tends to zero, such that scattering in low mass haloes (with typical velocities less than $v_\mathrm{max}$) is enhanced above the constant cross-section case. For $v_\mathrm{pair} \gg v_\mathrm{max}$, $\sigma_T \propto v_\mathrm{pair}^{-4}$, leading to a strong suppression of the scattering rate in DM haloes with velocity dispersions larger than $v_\mathrm{max}$.

The vdSIDMa and vdSIDMb models have values of $\sigma_T^{\rm max}/m$ and $v_\mathrm{max}$ chosen to maximise the self-interaction rate at the typical velocity dispersion of Milky Way dwarf spheroidals, while avoiding known astrophysical constraints on the cross-section. Specifically, vdSIDMa and vdSIDMb have $v_\mathrm{max} = 30 \, \mathrm{km \, s^{-1}}$ and $\sigma_T^{\rm max}/m = 3.5 \, \mathrm{cm^2 \, g^{-1}}$, and  $v_\mathrm{max} = 10 \, \mathrm{km \, s^{-1}}$ and $\sigma_T^{\rm max}/m = 35 \, \mathrm{cm^2 \, g^{-1}}$ respectively.

\subsection{vdSIDM cosmic scattering rates}

The calculation of the DM scattering rate $\Gamma(z)$ proceeds in a similar manner to Section \ref{sect:Cosmic_Scatter_rate}, in that we first find the distribution of haloes of different mass ($\frac{\rmd F}{\rmd \, \ln M}$) and then find the scattering rate per unit mass in these haloes, $\Gamma_\mathrm{halo}(M)$. However, the calculation of $\Gamma_\mathrm{halo}(M)$ is complicated by the velocity-dependent cross-section, because the cross-section can no longer be taken outside the integral in Equation \eqref{Gamma_halo}. Instead, we find $\langle \sigma\, v_\mathrm{pair}\rangle(r)$ by numerically integrating $\sigma_T(v_\mathrm{pair}) \, v_\mathrm{pair}$ over the probability distribution of pairwise velocities, again assuming that the velocities of individual particles are drawn from a Maxwell-Boltzmann distribution function with 1D velocity dispersion $\sigma_{\rm 1D}$. This yields
\begin{equation}
\left<\sigma_T v\right>(\sigma_{\rm 1D})=\frac{1}{2\sigma_{\rm 1D}^3\sqrt{\pi}}\int \sigma_T(v) v^3 e^{-v^2/4\sigma_{\rm 1D}^2}\,{\rm d}v.
\end{equation}
Then with $\sigma_{\rm 1D}(r)$ from Equation \eqref{eq:v_disp} we can find $\left<\sigma_T v\right>(r)$, which we use in the numerical evaluation of Equation \eqref{Gamma_halo} to calculate $\Gamma_\mathrm{halo}(M)$. Combining $\Gamma_\mathrm{halo}(M)$ with the multiplicity function we can calculate the cosmic scattering rate as in Section \ref{sect:DM_cosmic_interaction_rate}.

\begin{figure*}
        \centering
        \includegraphics[width=\textwidth]{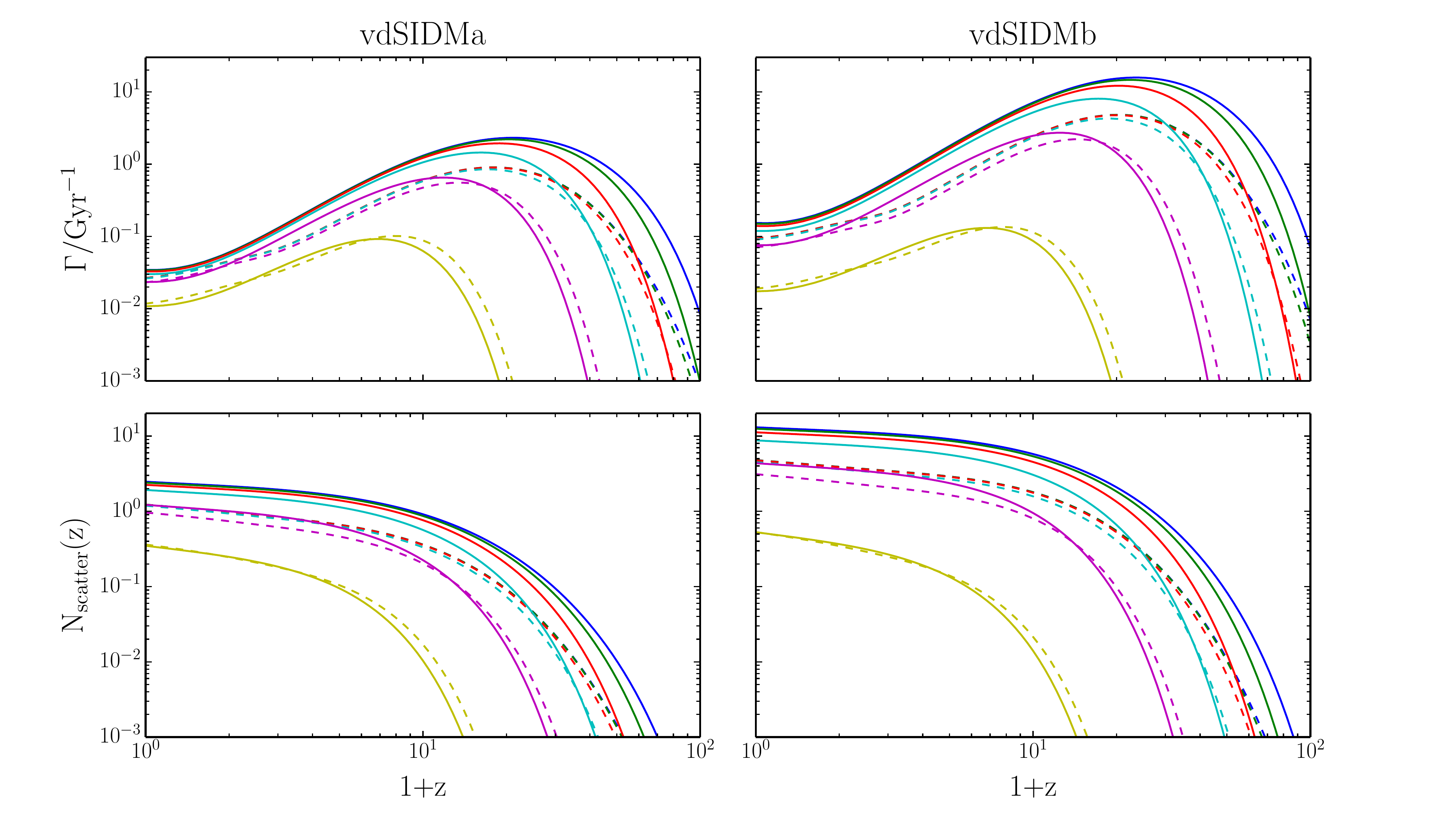}
	\caption{
Scattering rates (top row) and cumulative number of scatters (bottom row) as a function of redshift, for two different velocity-dependent scattering cross-sections. The left column is for vdSIDMa which has $v_\mathrm{max} = 30 \, \mathrm{km \, s^{-1}}$ and $\sigma_\mathrm{max}/m = 3.5 \, \mathrm{cm^2 \, g^{-1}}$; while vdSIDMb (right column) has $v_\mathrm{max} = 10 \, \mathrm{km \, s^{-1}}$ and $\sigma_\mathrm{max}/m = 35 \, \mathrm{cm^2 \, g^{-1}}$. The different line colours correspond to different values for $M_\mathrm{min}$ of $10^{8}, 10^{4}, 1, 10^{-4}, 10^{-8},$ and $10^{-12} \, \rmm_\odot$, with both $\Gamma$ and $N_\mathrm{scatter}$ monotonically increasing with decreasing $M_\mathrm{min}$. The solid lines are for the D08 concentration-mass-redshift relation, while the dashed lines use the P12 $c(M,z)$. \emph{Unlike the constant cross-section case in Fig. \ref{fig:cosmic_scatter_rate}, $\Gamma(z)$ is now plotted on a logarithmic scale as the scattering rate is larger by around two orders of magnitude at high redshift compared to redshift zero.}
	}
	\label{fig:vdSIDM}
\end{figure*}

\begin{figure*} 
\centering
\begin{minipage}[t]{\columnwidth}
\centering
        \includegraphics[width=\columnwidth]{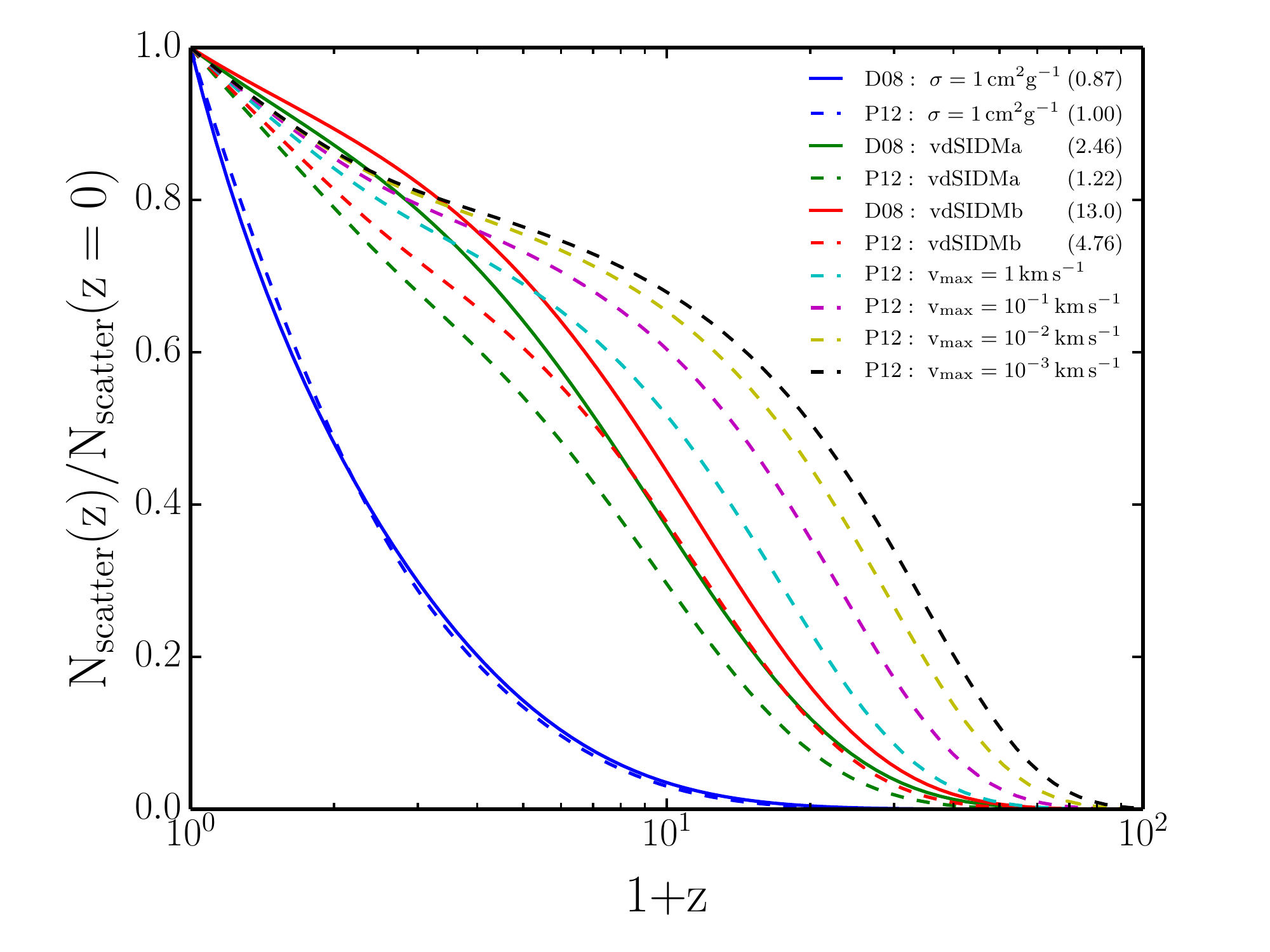}
	\caption{\emph{When do scatterings happen?} The cumulative number of interactions as a function of redshift, normalised to unity at redshift zero. The different colours correspond to different particle models for the DM, while the solid and dashed lines are for the D08 and P12 $c(M,z)$ relations respectively. All curves were calculated using $M_\mathrm{min} = 10^{-12} \, \rmm_\odot$. The number in brackets in the legend is $N_\mathrm{scatter}(z=0)$ for the relevant model. These are not present for the models with specified $v_\mathrm{max}$, which represent vdSIDM models with unspecified $\sigma_T^\mathrm{max}$. Velocity-dependent models with low $v_\mathrm{max}$ lead to more interactions in haloes with low internal velocities, pushing scattering towards high redshifts where collapsed objects are less massive.}
	\label{fig:cumm_scats_all_normalised}
\end{minipage}\hfill
\begin{minipage}[t]{\columnwidth}
\centering
        \includegraphics[width=\columnwidth]{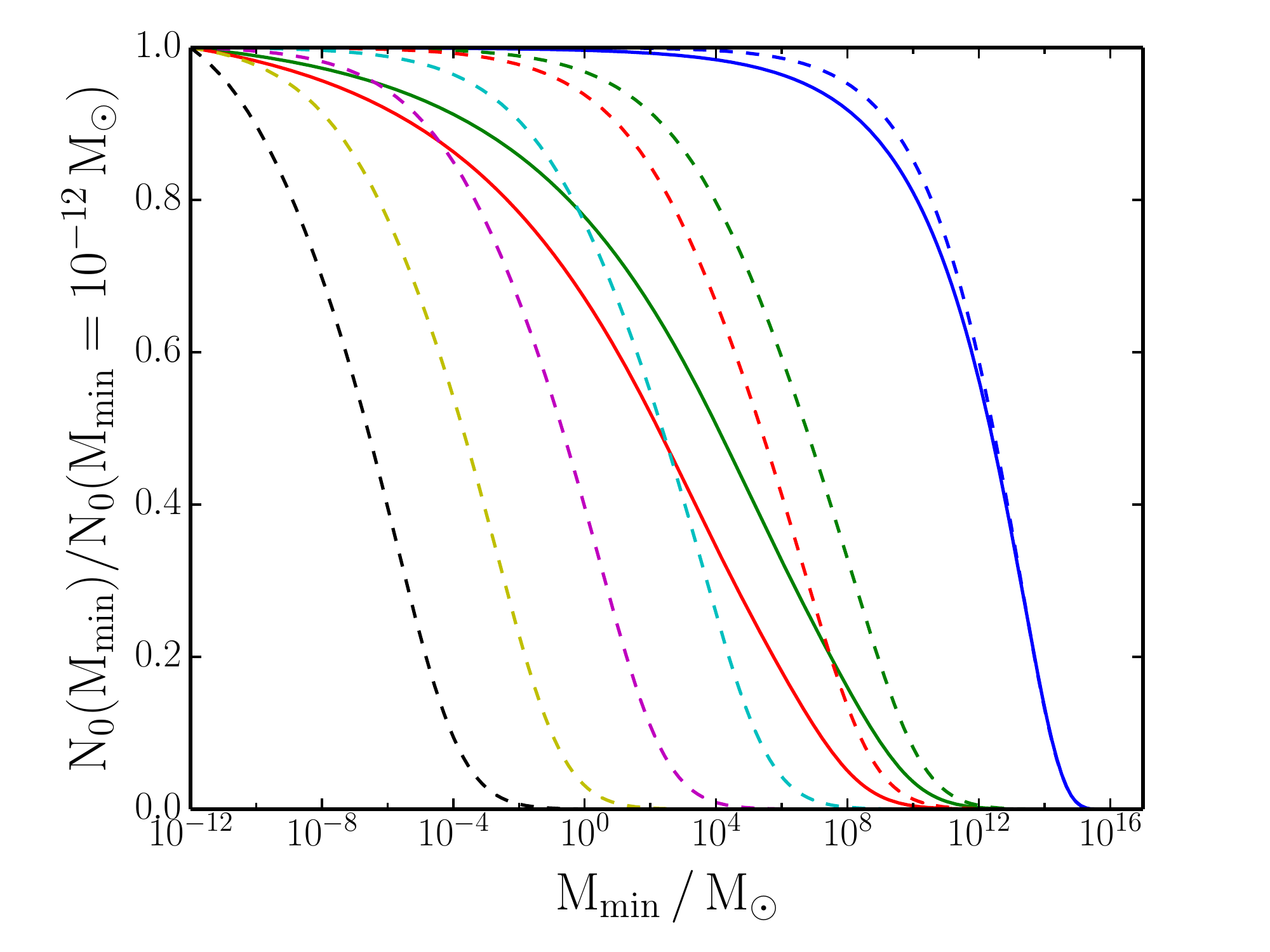}
	\caption{\emph{Where do scatterings happen?} The fraction of scatterings by redshift zero that occur in haloes more massive than $M_\mathrm{min}$, normalised to unity for $M_\mathrm{min} = 10^{-12} \, \rmm_\odot$. Different line styles are as in Fig. \ref{fig:cumm_scats_all_normalised}, with colours corresponding to particle models, and solid or dashed lines representing the D08 or P12 concentration-mass-redshift relations respectively. Models with velocity-independent cross-sections have more of their scatterings in high-mass haloes compared to velocity-dependent cases, where the typical halo mass in which most interactions happen is an increasing function of $v_\mathrm{max}$.}
	\label{fig:Mmin}
\end{minipage}
\end{figure*}

The scattering rate through cosmic time is plotted for vdSIDMa and vdSIDMb in Fig.~\ref{fig:vdSIDM}. In contrast to the velocity-independent case in Fig.~\ref{fig:cosmic_scatter_rate}, the scattering rate is now displayed on a logarithmic scale. It peaks at redshift $20-30$ and falls by two orders of magnitude before $z=0$. 
Most interactions thus occur at early times as can be seen in Fig.~\ref{fig:cumm_scats_all_normalised}.
Half occur before $z=5.7$ for vdSIDMa and $z=7.2$ for vdSIDMb (in the latter case, the Universe is $\sim 5\%$ of its present age). 
This is in stark contrast to the gentler evolution of $\Gamma(z)$ with a constant cross-section (c.f.\,Fig.~\ref{fig:cosmic_scatter_rate}), where half the interactions occur after $z=0.96$.

With a velocity-dependent cross-section, most scatterings also occur in low mass haloes with typical velocities $v \lesssim v_\mathrm{max}$.
Raising the minimum mass of considered haloes $M_\mathrm{min}$ from $10^{-12} \, \rmm_\odot$ to $10^{8} \, \rmm_\odot$ lowers the number of interactions by redshift zero by a factor of six, which can be seen in Fig.~\ref{fig:Mmin} (for which we introduce $N_0 \equiv N_\mathrm{scatter}(z=0)$). 
For the constant cross-section case, the same change leads to a decrease in $N_\mathrm{scatter}(z=0)$ of only 10\%.

The choice of concentration-mass-redshift relation becomes more important when the cross-section is velocity-dependent, because different $c(M,z)$ disagree most for low mass haloes and at high redshift. In particular, the simple power law relation from D08 predicts low-mass haloes to be much more concentrated than more recent relations in which $c(M)$ flattens at low mass. This recovers (a less extreme version of) what is seen in estimates of the DM annihilation rate, where $\langle \sigma\, v_\mathrm{pair}\rangle$ is usually assumed to be constant, resulting in an even larger fraction of interactions occurring in low mass haloes, and hence a cosmic scattering rate with strong dependence on $c(M,z)$ \citep{2014MNRAS.439.2728M,2015MNRAS.452.1217C}. 
 
As well as the three particle models already discussed (velocity-independent, vdSIDMa and vdSIDMb), we include in Fig.s \ref{fig:cumm_scats_all_normalised} and \ref{fig:Mmin} plausible but more extreme velocity-dependent models with lower $v_\mathrm{max}$.
We need not specify the normalisation of $\sigma_T^{\rm max}/m$ for these calculations, but it can be chosen to solve small-scale problems at dwarf galaxy scales, while eluding constraints at cluster scales.
As $v_\mathrm{max}$ is lowered, a larger fraction of interactions happen at high redshift and in low-mass haloes. 
For the most extreme case considered, with $v_\mathrm{max}=10^{-3} \, \mathrm{km \, s^{-1}}$, half of the interactions have occurred by $z=19$, and half occur in haloes of mass $<10^{-6} \, \rmm_\odot$. We stress that such models cannot be excluded on particle physics grounds, but it is unclear whether the large number of scatterings in such low mass haloes would leave a detectable signal in the present day universe.

\section{Conclusions}
\label{sect:conclusions}

We have presented an analytical calculation of the mean rate of dark matter-dark matter scattering events,
for particle physics models with a velocity-independent or velocity-dependent cross-section. 
In all our calculations, we assume that the self-interactions are a small perturbation to $\Lambda$CDM and do not, for example, change the overall growth of structure.

For particle physics models with a velocity-independent interaction cross-section, our results match the canonical picture in which most scatterings occur in massive structures $\simgt10^{12} \, \rmm_\odot$ at late times $z\simlt1$.
Our calculations are found to be robust to current uncertainties in cosmological parameters as well as variations in the mass function used. They are also insensitive to the high-$k$ power spectrum (because most scattering events occur in haloes more massive than the cut-off scales due to DM self-interactions in the early Universe).
The main source of uncertainty in the results is the concentration-mass-redshift relation $c(M,z)$. Its unknown form at high redshift and low mass propagates into a factor of almost three discrepancy in the scattering rate at intermediate redshifts ($z\approx10$). However, the scattering rate changes by only a factor of two over most of cosmic time, and different $c(M,z)$ relations give similar results after $z \approx 1$, where there is more time. Consequently, the total number of interactions during the entire history of the Universe is uncertain to only a factor of $\sim2$.

For particle physics models with a well-motivated velocity dependence, the scattering takes place mainly in low mass objects $\simlt10^{4} \, \rmm_\odot$ at early times $z\simgt7$.
The scattering rate $\Gamma (z)$ peaks at earlier redshifts $z\sim20$, with a pronounced peak two orders of magnitude higher than the scattering rate at the present day.
These numbers are more sensitive to the choice of cosmological and astrophysical parameters, and are dominated by regimes in which the mass function and concentration-mass-redshift relation are least well known.
This minimum mass of considered structures, $M_\mathrm{min}$, is particularly important, with changes to the small-scale power spectrum induced by DM scattering affecting the cosmic scattering rate. 

The dominance of high redshift scatterings in velocity-dependent models profoundly changes their influence on the evolution of structure, and may alter the best strategy to search for observational signatures.
DM particle interactions lead to a transport of particles away from the centres of structures \citep{2000ApJ...543..514K}, replacing the cusps found in collisionless CDM simulations with constant density cores.\footnote{Haloes in which there have been a larger number of interactions presumably have larger cores. However, we caution against qualitative attempts to determine the scattering rate from core sizes: estimates of the core size and ellipticity of a galaxy cluster halo \citep{2002ApJ...564...60M} overestimated the effect of SIDM by a factor $50$ compared to full simulations \citep{2013MNRAS.430..105P}.
Furthermore, strong gravitational lensing measurements of the cores in low redshift clusters \citep{2003AAS...203.3004S,2013ApJ...765...24N} are subject to projection effects. Particularly when SIDM lowers the central density, material at large radii significantly contributed to the 2D projected density.}
If the SIDM interactions are effectively confined to high redshift, then they may lead to a smearing of small-scale structure more qualitatively reminiscent of warm dark matter.
The affected DM structures are also the hosts of the first galaxies, and it is interesting to consider what impact cored haloes could have on early galaxy formation.

High redshift scattering in low-mass objects also has important consequences for attempts to simulate vdSIDM cosmologies. Most scatterings occur in low mass haloes at high redshift that would not be resolved in typical cosmological simulations, but the unresolved interactions could be important for the later dynamics of particles. The large number of self-interactions would create DM cores in high-redshift haloes, and it then becomes an important question -- on which there seems little consensus -- whether or not the mergers of small cored haloes form cores that persist in large haloes at the present day.

\section*{Acknowledgments}

This work was supported by the Science and Technology Facilities Council grant numbers ST/K501979/1 and ST/L00075X/1. RM was supported by the Royal Society.

\bibliographystyle{mnras}

\bibliography{MS}

\bsp
\label{lastpage}

\end{document}